\documentclass[twocolumn]{aastex62}
\usepackage{graphicx}
\usepackage{color}
\usepackage{amsmath}
\usepackage{bm}
\usepackage{comment}
\usepackage{lineno}
\usepackage[first=0,last=9]{lcg}

\usepackage{array,multirow}

\definecolor{LightCyan}{rgb}{0.88,1,1}

\usepackage{appendix}
\usepackage{tabularx}
\usepackage{enumerate}
\usepackage{enumitem}

\newcommand\unit[1]{{\rm #1}}

\newcommand\qmstateproduct[2]{\left<#1|#2\right>}

\usepackage{subfig}
\usepackage[normalem]{ulem}

\begin{document}

\title{\bf  Eccentricity estimation for five binary black hole mergers with higher-order gravitational wave modes}

\author{H. L. Iglesias}
\affiliation{Center of Gravitational Physics, University of Texas at Austin, Austin, TX 78712, USA}
\thanks{higlesia@utexas.edu}
\author{J. Lange}
\thanks{jacob.lange@austin.utexas.edu}
\affiliation{Center of Gravitational Physics, University of Texas at Austin, Austin, TX 78712, USA}
\author{I. Bartos}
\thanks{imrebartos@ufl.edu}
\affiliation{Department of Physics, University of Florida, PO Box 118440, Gainesville, FL 32611-8440, USA}
\author{S. Bhaumik}
\affiliation{Department of Physics, University of Florida, PO Box 118440, Gainesville, FL 32611-8440, USA}
\author{R. Gamba}
\affiliation{Theoretisch-Physikalisches Institut, Friedrich-Schiller-Universit {\"a}t Jena, 07743, Jena, Germany}
\author{V. Gayathri}
\affiliation{Department of Physics, University of Florida, PO Box 118440, Gainesville, FL 32611-8440, USA}
\affiliation{Leonard E. Parker Center for Gravitation, Cosmology, and Astrophysics, University of Wisconsin–Milwaukee, Milwaukee, WI 53201, USA}
\author{A. Jan}
\affiliation{Center of Gravitational Physics, University of Texas at Austin, Austin, TX 78712, USA}
\author{R. Nowicki}
\affiliation{Center of Gravitational Physics, University of Texas at Austin, Austin, TX 78712, USA}
\author{R. O'Shaughnessy}
\affiliation{Rochester Institute of Technology, Rochester, NY 14623, USA}
\author{D. M. Shoemaker}
\affiliation{Center of Gravitational Physics, University of Texas at Austin, Austin, TX 78712, USA}
\author{R. Venkataramanan}
\affiliation{Center of Gravitational Physics, University of Texas at Austin, Austin, TX 78712, USA}
\author{K. Wagner}
\affiliation{Rochester Institute of Technology, Rochester, NY 14623, USA}

\begin{abstract}
The detection of orbital eccentricity for a binary black hole system via  gravitational waves is a key signature to distinguish between the possible binary origins. The identification of eccentricity has been difficult so far due to the limited availability of eccentric gravitational waveforms over the full range of black hole masses and eccentricities. Here we evaluate the eccentricity of five black hole mergers detected by the LIGO and Virgo observatories for the first time using the {\tt TEOBResumSGeneral} model. This model accounts for the full eccentricity range possible and incorporates higher-order gravitational wave modes critical to model emission from highly eccentric orbits. The binaries have been selected due to previous hints of eccentricity or due to their unusual mass and spin. While other studies found marginal evidence for eccentricity for some of these events, our analyses do not favor the incorporation of eccentricity compared to the quasi-circular case. While lacking the eccentric evidence of other analyses, we find our analyses marginally shifts the posterior in multiple parameters for several events when allowing eccentricity to be non-zero.
\end{abstract}


\section{Introduction}
\label{sec:intro}
The LIGO \citep{2015CQGra..32g4001L} and Virgo \citep{2015CQGra..32b4001A} observatories have discovered about 90 binary mergers via gravitational waves (GWs) signals over three observing periods \citep{2018arXiv181112907T,2020arXiv201014527A,LIGOScientific:2021djp}. The properties of these mergers have been estimated using gravitational wave templates that assume quasi-circular binary orbits prior to merger \citep{2020PhRvD.102d4055O,2019PhRvD.100b4059K, 2021PhRvD.103j4056P}. This approach has been the baseline due to both prior expectations and technical challenges. Gravitational wave emission is expected to circularize binaries over time, making eccentricity unexpected for the majority of binaries. At the same time the construction of accurate eccentric gravitational waveforms is difficult due to the required large dynamic range, while accounting for eccentricity also requires probing an expended parameter space. Nonetheless, searches carried out by LIGO and Virgo are sensitive to eccentric binaries \citep{Abbott_eBBH}, leaving the possibility that some of the recovered mergers are secretly eccentric.

While long-lived, isolated binaries lose their eccentricity before entering the LIGO/Virgo frequency band, measurable eccentricity is expected through several astrophysical mechanisms. First, binary black hole systems that form with small initial separations may not have sufficient time prior to merger to lose their eccentricity. This scenario can occur, e.g. in chance encounters in galactic nuclei or globular clusters \citep{2000ApJ...528L..17P,2010MNRAS.402..371B,2014MNRAS.441.3703Z,2015ApJ...800....9M,2016MNRAS.459.3432M,2016PhRvD..93h4029R,2017MNRAS.464L..36A}. Eccentricity can also be enhanced for long-lived binaries by a nearby third object via the Kozai-Lidov mechanism \citep{2016ARA&A..54..441N,2017ApJ...841...77A,2018ApJ...853...93R}. In active galactic nuclei (AGN), binaries formed via gas capture in AGN disks can have enhanced eccentricity due to regular encounters with stars and black holes in the galactic nucleus (\citealt{2022Natur.603..237S,2021ApJ...907L..20T,2017ApJ...835..165B}).

Recently, several techniques have been developed and applied to probe orbital eccentricity in binary black hole mergers detected by LIGO and Virgo. \cite{2022NatAs...6..344G} used a bank of about 600 numerical relativity simulations throughout the full eccentricity range and estimated their match with the observed data for the binary merger GW190521 \citep{GW190521discovery} using the RIFT parameter estimation suite \citep{gwastro-PENR-RIFT,gwastro-PENR-Methods-Lange,dissertation-RIFT-Lange}. They found that the observed data best matches a highly eccentric waveform, and reconstructed the eccentricity of GW190521 to be $e=0.69^{+0.17}_{-0.22}$. While applicable to GW190521, their technique was limited to binaries with high masses due to the duration of the simulated waveforms, while the reconstruction's precision was limited by the limited number of discrete eccentricity, spin and mass ratio values covered in the study. 

\cite{IsobelGW190521} used gravitational wave template banks that covered the full mass parameter space, while limited to eccentricities $e<0.2$ and black hole spins aligned with the binary orbit. They performed their inference by a novel postprocessing technique: they first analyze with the quasi-circular model IMRPhenomD \citep{2011PhRvL.106x1101A} and, based on those results, reweight their conclusions with the aligned-spin eccentric model SEOBNRE \citep{2019PhRvD.100l3017P,2019MNRAS.490.5210R}.  A further possible limitation of this approach was the iterative reweighting which may have been limited by systematic errors between the two models used in the reweighted analysis \citep{2021ApJ...921L..31R,2020PhRvD.102l4069J}. 

\cite{2021arXiv210707981O} and \cite{2021arXiv210605575G} performed parameter inference including the continuously-explored effects of eccentricity and assuming black hole spins aligned with the orbital axis (non-precessing spins). \cite{2021arXiv210707981O} analyzed the low-mass events GW151226 and GW170608 including eccentricity and non-precessing spins; \cite{2021arXiv210605575G} analyzed GW190521 with non-spinning systems and sampled in initial energy and angular momentum instead of eccentricity which could impact the prior volume and therefore the evidence. These analytic models assumed spin-orbit alignment, limiting attempts to disentangle the effects of orbital eccentricity and precession \citep{PhysRevLett.126.201101}.

Currently, there are two easily-accessible waveform models which predict the strong-field merger signal for merging binary black holes which incorporate both binary spin and orbital eccentricity: SEOBNRE and TEOBResumSGeneral. The SEOBNRE model is an effective-one-body (EOB) \citep{Buonanno:1998gg,Buonanno:2000ef,Damour:2000we,Damour:2001tu,Damour:2015isa} model that is based on the early quasi-circular nonprecessing model SEOBNRv1. While this model can describe eccentric binary black hole mergers, it is limited to eccentricities $e<0.2$ and spin magnitudes $a<0.6$ due to being based on the older v1 SEOB model (\cite{2017PhRvD..96d4028C}). 
The {\tt TEOBResumSGeneral} model is an EOB approximant which can generate waveforms from non-precessing binaries coalescing along generic orbits \citep{Chiaramello:2020ehz, Nagar:2020xsk, 2021PhRvD.103j4021N}, or generic-spins binaries coalescing along quasi-circular orbits \citep{Damour:2014sva, Nagar:2015xqa, Nagar:2018zoe, Nagar:2019wds, Nagar:2020pcj, Riemenschneider:2021ppj, Akcay:2020qrj, Gamba:2021ydi}. In both scenarios, the model includes tidal interactions \citep{Damour:2009wj, Bernuzzi:2014owa, Akcay:2018yyh} as well as subdominant modes up to $\ell=|m|=5$ but not $m=0$ modes. While the model can generate stable waveforms up to an eccentricity of $e=0.9$ with maximal spin magnitudes, comparisons to numerical relativity (NR) waveforms have only been carried out up to eccentricities of $e=0.2$ and spin magnitudes up to $a=0.7$ \citep{2021PhRvD.103j4021N}. Another EOB model in the literature that incorporates orbital eccentricity is SEOBNRv4EHM, which also describes non-precessing spins and includes higher order modes \citep{2021PhRvD.104b4046K,2022PhRvD.105d4035R}.

In this study, we analyzed a group of GW events to produce full posteriors using the eccentric model TEOBResumSGeneral in tandem with parameter estimation carried out using RIFT. Events were selected based on signs of possible eccentricity in previous studies, or due to their unusual mass/spin properties. For each event, we run \textbf{three} different analyses (\textbf{two} show in the figures): a non-precessing, non-eccentric analysis (from now on called TEOBResumS-GIOTTO); a precessing, non-eccentric analysis (from now on called TEOBResumSP-GIOTTO); and a non-precessing, eccentric analysis (from now on called TEOBResumS-DALI).
To quantify any evidence in these analyses, we calculate Bayes' factors for eccentricity as well as precession. We include the latter to make sure any evidence of eccentricity does not also yield evidence of precession. This could potentially happen if an event is in a region of parameter space where precession and eccentricity can mimic each other in the analysis~\citep{2023MNRAS.519.5352R}.

This paper is organized as follows. In Section \ref{sec:methods}, we introduce the TEOBResumSGeneral model and review the use of RIFT in this study. In Section \ref{sec:results}, we present the results of parameter inference as well as Bayes' factors for our five events. In Section \ref{sec:conclusions}, we summarize our results and conclude with some brief remarks about our future work.

\section{Methods}
\label{sec:methods}

A coalescing binary black hole system in a quasi-circular orbit can be completely characterized by its intrinsic ($\lambda$)
and extrinsic ($\theta$) parameters.  By intrinsic parameters we refer to the binary's  masses $m_i$ and spins.
By extrinsic parameters we refer to the seven numbers needed to characterize its space-time location and orientation.  
We will express masses in solar mass units and
 dimensionless spins in terms of cartesian components $\chi_{i,x},\chi_{i,y}, \chi_{i,z}$, expressed
relative to a frame with $\hat{\mathbf{z}}=\hat{\mathbf{L}}$ and (for simplicity) at the orbital frequency corresponding to the earliest
time of computational interest (e.g., an orbital frequency of $\simeq 10 \,\unit{Hz}$).  We will use $\lambda,\theta$ to
refer to intrinsic and extrinsic parameters, respectively.

\subsection{Waveform model}
\label{subsec:model}
In order to extract the eccentricity feature from detected GW events, we are using the TEOBResumSGeneral waveform model \citep{Nagar:2021gss,Nagar:2018zoe,Albanesi:2022xge}, which includes eccentricity features, aligned spin, and higher order modes. This waveform family is based on the EOB formalism and it can be used to simulate quasi-circular, eccentric or hyperbolic compact binaries systems. This model includes higher multipoles up to 5  throughout binary phases (inspiral, plunge, merger and ringdown) and eccentricity parameter space (i.e., up to $e\simeq 1$; see figure 19 in \cite{Nagar:2021gss}).  

TEOBResumSGeneral waveforms have been validated with Numerical Relativity waveforms from the Simulating eXtreme Spacetime collaboration with $e\leq0.2$, and mass ratio $q\leq3$. For our study we used TEOBResumSGeneral waveforms with three different configurations, corresponding to  $e=0$ with non-precessing spins (TEOBResumS-GIOTTO), $e=0$ with precessing spins (TEOBResumSP-GIOTTO), and $e>0$ with non-precessing spins (TEOBResumS-DALI).

To generate orbital dynamics from eccentric initial conditions -- an initial eccentricity $e_0$ at initial orbital frequency $f_0$ -- the TEOBResumS-DALI code evolves an orbit starting at apastron, with $r_0=p_0/(1-e_0)$, $p^0_\phi=j_0$ the adiabatic angular  momentum implied by angular momentum conservation, and $p_{r_*}=0$.  In these expressions, $p_0$ is evaluated numerically from the Hamiltonian such that $\partial_{p_\phi} H( r_0(p_0),j_0(p_0), p_{r_*}=0)=f_0$.   
The orbital dynamics produce an associated asymptotic gravitational wave strain $h(t,\hat{n}) = \sum_{lm}  {}^{-2}Y_{lm}(\hat{n}) h_{lm}(t)$, where the $h_{lm}(t)$ depend on the intrinsic parameters.  In practice, we characterize the emission direction relative to the orbital angular momentum direction with two polar angles ($\iota,\phi_{c}$), so the binary's radiation relative to its fiducial inertial frame is fully specified with an initial frequency (to define conventions), the initial eccentricity at that frequency, two masses,  both spins, and these two polar angles.
While this current parameterization fixes the angle of periapsis (one degree of freedom available for eccentric orbits associated with the direction of the Runge-Lenz vector in the orbital plane), this angle is expected to be observationally inaccessible in the near future \citep[see, e.g.,][]{2022arXiv220614006C}.

\subsection{RIFT}
\label{subsec:rift}

RIFT  consists of a two-stage iterative process to estimate the source parameters $\bm \theta,\lambda$ responsible for gravitational wave observations $d_k$ via comparison to
predicted gravitational wave signals $h_k(\bm{\lambda}, \bm\theta)$ where  $k=1\ldots N$ indexes the observing instruments.

Assuming a Gaussian, stationary noise model, the log-likelihood  can be expressed as
\begin{align} \label{eq:loglikelihood}
\ln & {\cal L}(\lambda, \theta) = \nonumber \\
& -\frac{1}{2}\sum_k \qmstateproduct{h_k(\lambda,\theta)-d_k}{h_k(\lambda,\theta)-d_k}_k  
 - \qmstateproduct{d_k}{d_k}_k
\end{align}
where we have omitted normalization constants.  (RIFT assumes the input detector noise power spectrum is known, and does not currently marginalize over the accuracy of that estimate, nor over calibration uncertainty.)

In these expressions, the inner products represent the noise-weighted inner product derived from the $k$th detector's noise power spectrum
\begin{align*}
\qmstateproduct{a}{b}_k \equiv \int_{-\infty}^{\infty} 2 df \frac{\tilde{a}(f)^*\tilde{b}(f)}{S_{n,k}(|f|)} \,,
\end{align*}
is a 
$S_{n,k}(f)$, where $\tilde{a}(f)$ is the Fourier transform of $a(t)$,
$\tilde{a}(f)^*$ denotes complex conjugation of $\tilde{a}(f)$,
and $f$ is frequency;  see, e.g., \cite{gwastro-PE-AlternativeArchitectures} for more details. 
We adopt a low-frequency cutoff $f_{\rm low}$
such that all inner products are modified to
\begin{eqnarray}
\qmstateproduct{a}{b}_k\equiv 2 \int_{|f|>f_{\rm low}}  df \frac{\tilde{a}(f)^*\tilde{b}(f)}{S_{n,k}(|f|)} \,.
\end{eqnarray}

The two iterative stages of RIFT construct the necessary ingredients for an iterative evaluation of Bayes' theorem for this likelihood.
In  one stage of RIFT, many worker codes evaluate the \emph{marginal} likelihood  $\ln {\cal L}_\alpha \equiv \ln {\cal L}_{\rm marg}(\lambda_\alpha)$ for $\lambda_\alpha$ in some grid of evaluation points, via
\begin{align}
{\cal L}_{\rm marg}(\lambda) \equiv \int d\theta p(\theta) {\cal L}(\lambda,\theta)
\end{align}
In the other iterative stage of RIFT, an interpolation algorithm provides a current-best-estimate $\hat{\cal L}(\lambda)$ based on current training data $\{(\lambda_\alpha,{\cal L}_\alpha\}$, and employs this estimate in Bayes' theorem to construct an approximate posterior distribution over the intrinsic parameters $\lambda$:
\begin{align}
    \hat{p}_{\rm post}(\lambda) \simeq  \frac{\hat{{\cal L}}(\lambda) p(\lambda)}{\int d\lambda \hat{{\cal L}}(\lambda) p(\lambda)}
\end{align}
Again using Monte Carlo integration, RIFT produces independent fair draws from this estimated posterior distribution, thus providing a new grid for the evaluation stage.
After several iterations, $\ln \hat{\cal L}$ will converge to the the true log-likelihood in the neighborhood where the posterior has its support, and the intrinsic samples will be the true posterior distribution.

\section{Results}
\label{sec:results}
In this section, we present the results from our eccentric reanalysis of the following events: GW150914, GW190521, GW190620, GW190706, GW190929. GW190521 and GW190620 were analyzed due to past evidence or hints of eccentricity (\cite{2022NatAs...6..344G, 2021ApJ...921L..31R}); GW150914, GW190706, and GW190929 were also picked due to specific characteristics (HOMs, unequal masses, high mass, positive spins, etc.). All the data and power spectral density (PSDs) are the same files used in the LVK's GWTC-2.1 paper (\cite{2021arXiv210801045T}) and are available on GWOSC \citep{ligo-O1O2-opendata}. Due to this, we chose 4 seconds of data around each event except for GW190521 where we chose 8 seconds of data\footnote{We chose to use 8 seconds of data instead of following GWTC-2.1 settings since the original analysis of this event was done with 8 seconds including \cite{2022NatAs...6..344G}.}. GW190521, GW190706, GW190929 uses data from all three HLV detectors with the HL data using the C01 calibration with noise subtracted out around 60 Hz (\cite{2021arXiv210801045T,2015PhRvD..91h4034L,2019PhRvD.100j4004C}). For these events, the V data uses the online calibration data except for GW190929 that uses 1A calibration. Finally GW150914 uses HL data using the C02 calibration, and GW190620 uses LV data using the C01 calibration with noise subtracted out around 60 Hz for L and using the online calibration for V. The PSDs were generated using BayesWave (\cite{2015PhRvD..91h4034L,2019PhRvD.100j4004C}) on the same segment of data used in each analysis.

For all events except GW190521, we used a low-frequency cutoff of 20 Hz, motivated by the lack of observationally-accessible information at lower frequencies for typical sources. For GW190521, we use a low-frequency cutoff of 11 Hz. The high-frequency cutoff was 896, 224, 448, 896, and 896 Hz for GW150914, GW190521, GW190620, GW190706, and GW190929 respectively. We analyzed each event with the non-precessing, eccentric TEOBResumS-DALI include all the $\ell_{\rm max}\le4$ except the $m=0$ modes. For comparison, we also analyzed each event with the non-precessing, non-eccentic TEOBResumS-GIOTTO including the same amount of modes as the eccentric analysis. An additional analysis with the precessing, non-eccentric TEOBResumSP-GIOTTO was conducted to ascertain evidence of precession in these events.

We adopted conventional masses and distance priors \citep[uniform in detector-frame mass and in the cube of the luminosity distance; see, e.g.,][]{gw-astro-PE-lalinference-v1}. For the precessing spins, we assume the spin vectors to be isotropic on the sphere and uniform in spin magnitude $\chi_{i,\{x,y,z\}}\in[-0.99,0.99]$.
For the non-precessing spins, we employ an aligned-spin prior called ``zprior," described in Appendix A of \cite{gwastro-PENR-RIFT}; we use this alternative prior to enable comparison to precessing spins as it is equivalent to the uniform spin magnitude prior after marginalizing out non-aligned spins.
For eccentricity, we adopted a uniform prior over the range $e \in [0.0, 0.6]$ for each event. Although the TEOBResumS-DALI model can generate stable waveforms with eccentricities up to $e \sim 0.9$, issues with the validity of these highly eccentric waveforms led to our choice to constrain the prior range; this is further explained in Appendix \ref{app:waveform model}.

\subsection{Masses}
\label{subsec:masses}
\begin{figure*}
    \subfloat{
        \includegraphics[width=.5\textwidth]{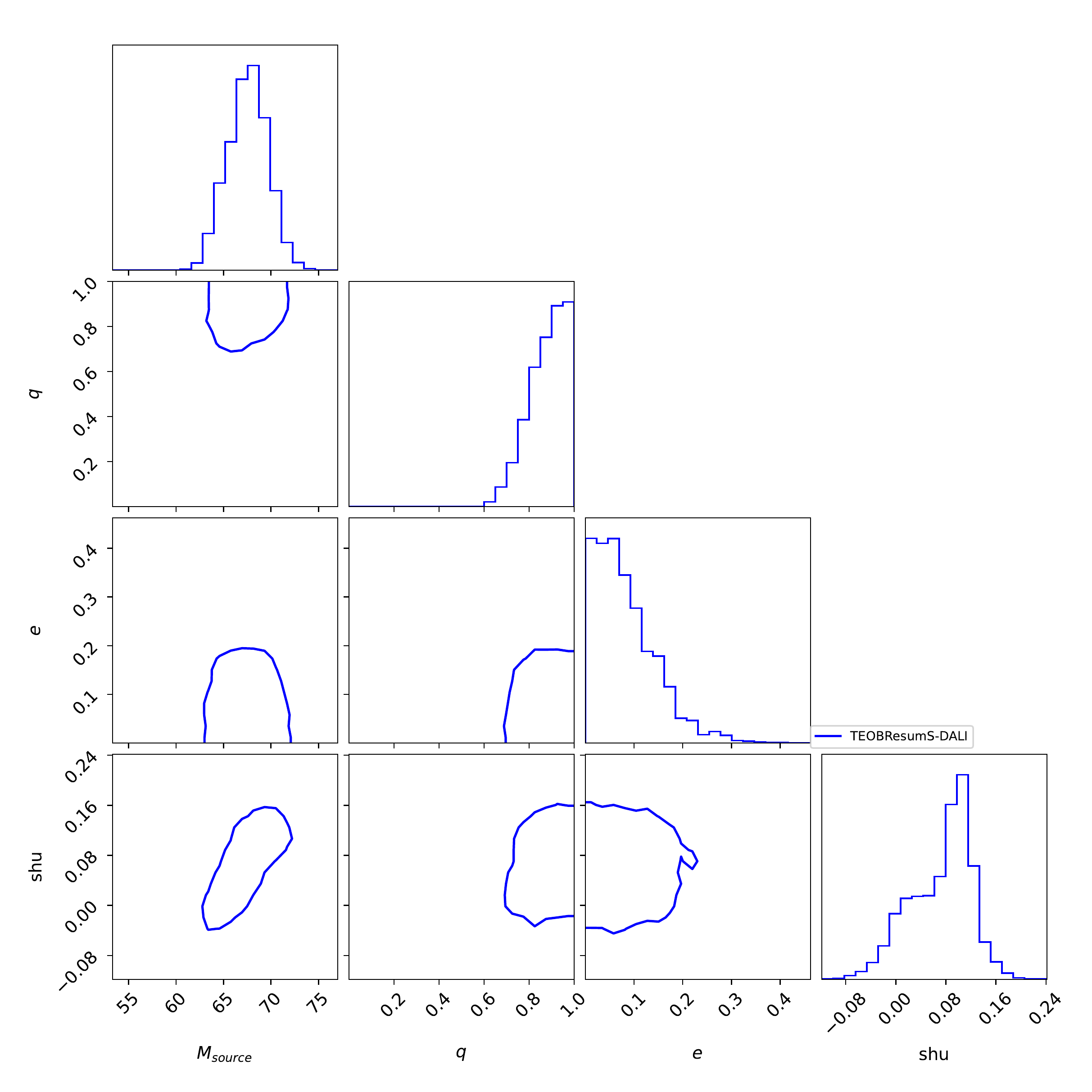}
    }
    \subfloat{
        \includegraphics[width=.5\textwidth]{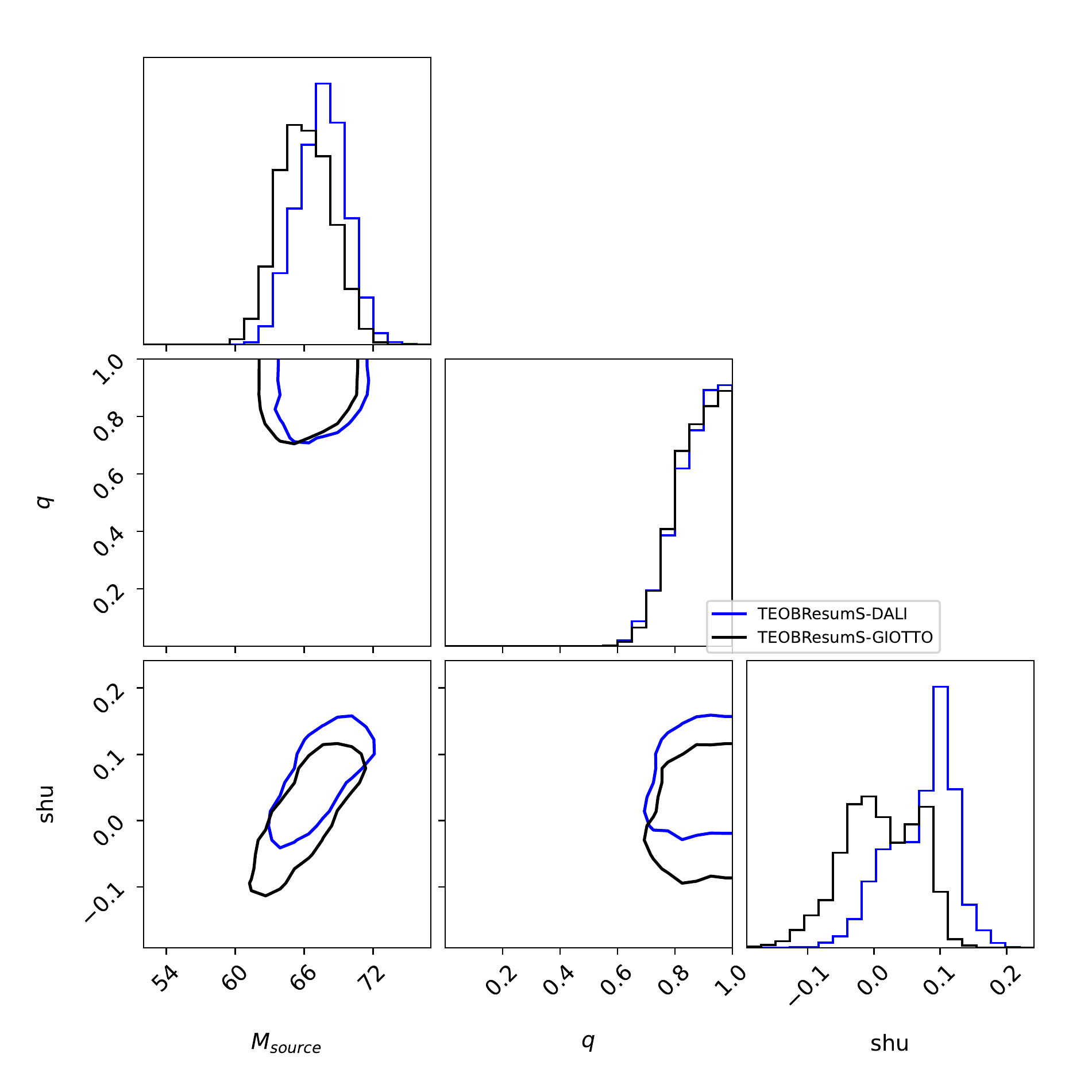}
    }
    \caption{
        \label{fig:GW150914}
        \textbf{GW150914}: Corner plots of 2D and 1D marginal posteriors of $M_{\rm source},q,e,S_{\rm hu}$ using TEOBResumS-GIOTTO in black and TEOBResumS-DALI in blue. \textit{The left panel} shows only the analysis using TEOBResumS-DALI and includes eccentricity in the corner plot along with masses and spin parameters. \textit{The right panel} includes both the TEOBResumS-DALI and TEOBResumS-GIOTTO analyses including only the masses and spin parameters. The contours represent the 90\% confidence intervals for each joint distribution. When eccentricity is included, it shifts the posterior of the masses and spins.
        }
\end{figure*}

\begin{figure*}
    \subfloat{
        \includegraphics[width=.5\textwidth]{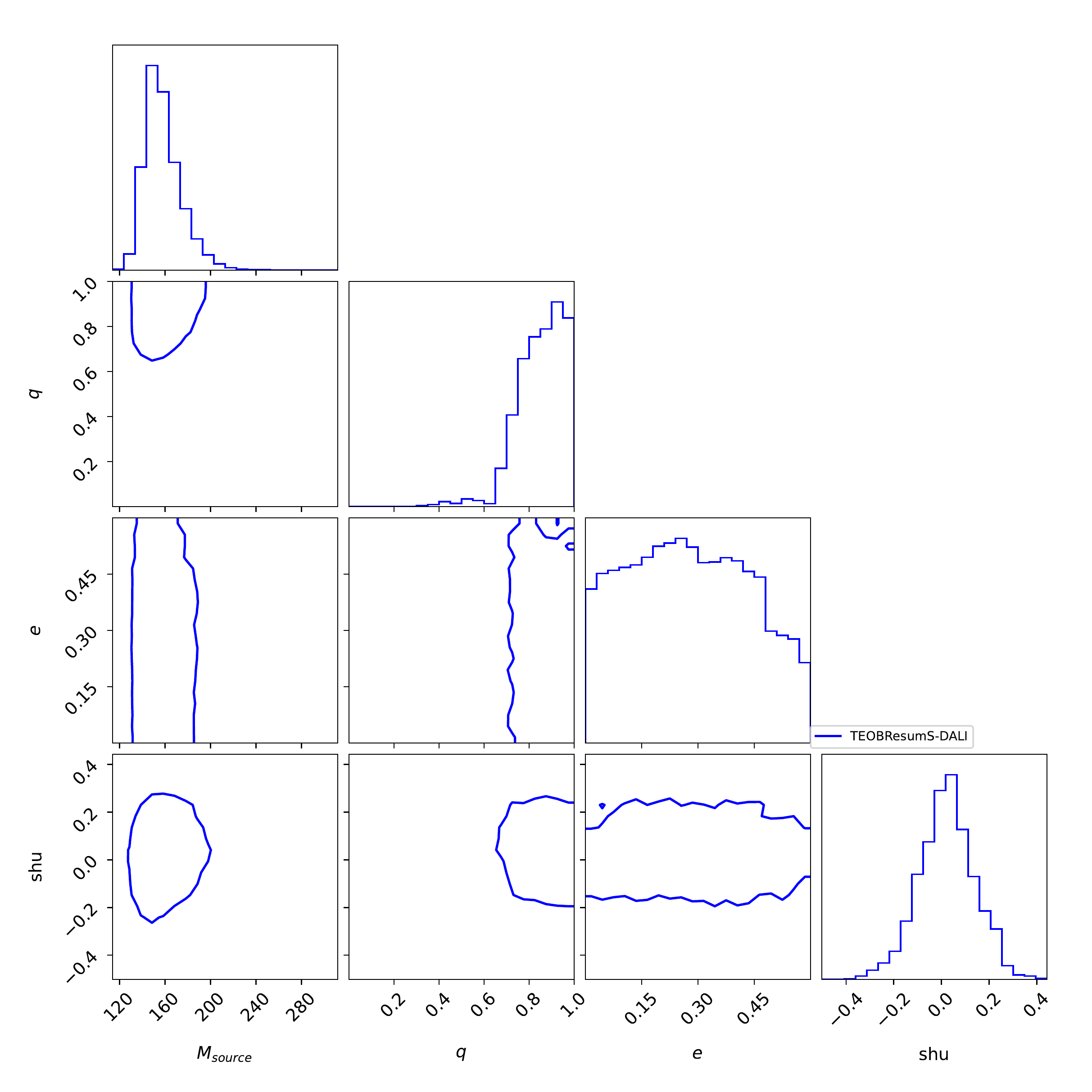}
    }
    \subfloat{
        \includegraphics[width=.5\textwidth]{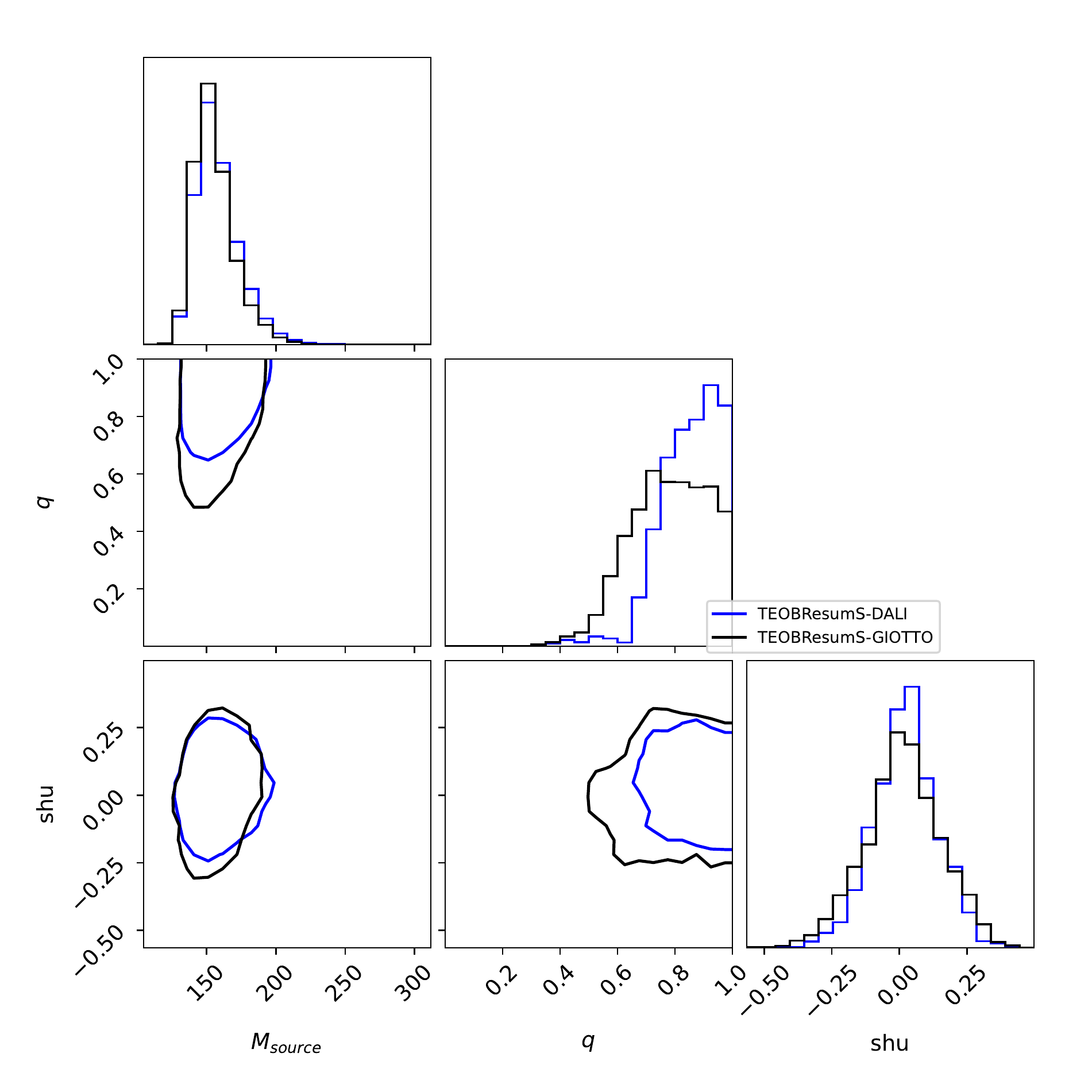}
    }
    \caption{
        \label{fig:GW190521}
        \textbf{GW190521}: Corner plots of 2D and 1D marginal posteriors of $M_{\rm source},q,e,S_{\rm hu}$ using TEOBResumS-GIOTTO in black and TEOBResumS-DALI in blue. \textit{The left panel} shows only the analysis using TEOBResumS-DALI and includes eccentricity in the corner plot along with masses and spin parameters. \textit{The right panel} includes both the TEOBResumS-DALI and TEOBResumS-GIOTTO analyses including only the masses and spin parameters. The contours represent the 90\% confidence intervals for each joint distribution. When eccentricity is included, it shifts the posterior of the mass ratio.
        }
\end{figure*}

Since we carried out analyses with both TEOBReumS-GIOTTO and TEOBResumS-DALI, we were able to make a direct comparison of the effect of the inclusion of eccentricity on each binary's component masses. For mosts events, the inferred mass parameters do not differ significantly depending on whether an analysis allowed for or excluded eccentricity. The only two examples of the mass parameters changing is GW150914 and GW190521. The right panels of Figures \ref{fig:GW150914} and \ref{fig:GW190521} shows the inferred 90\% credible intervals for the joint $M_{\rm source}-q$ distribution, derived from inference with both TEOBResumS-GIOTTO and TEOBResumS-DALI. In the GW150914 case, the $M_{\rm source}$ distributions are noticeably different with TEOBResumS-DALI slightly shifted toward larger masses while the $q$ distributions remain largely consistent. In GW190521 case, the $M_{\rm source}$ distributions are quite consistent in shape and support, while the $q$ distributions show modest differences in their shape and credible interval with TEOBResumS-GIOTTO shifted towards more unequal masses.

To quantify the relative change in parameter $x$ between the two analyses, we define $\epsilon_x = |\mu_1 - \mu_2|/\sqrt{\sigma_1^2+\sigma_2^2}$, where $\mu_k,\sigma_k$ are the posterior mean and standard deviation of $x$ for models $1$ (here, eccentric) and $2$ (here, non-eccentric) respectively. Between the events included in this study, we found the the largest relative shift for the total mass and mass ratio to be $\epsilon_M=0.472$ from GW150914 (See the right panel of Figure \ref{fig:GW150914}) and $\epsilon_q=0.432$ from GW190521 (See the right panel of Figure \ref{fig:GW190521}) respectively.

\begin{figure*}
    \subfloat{
        \includegraphics[width=.5\textwidth]{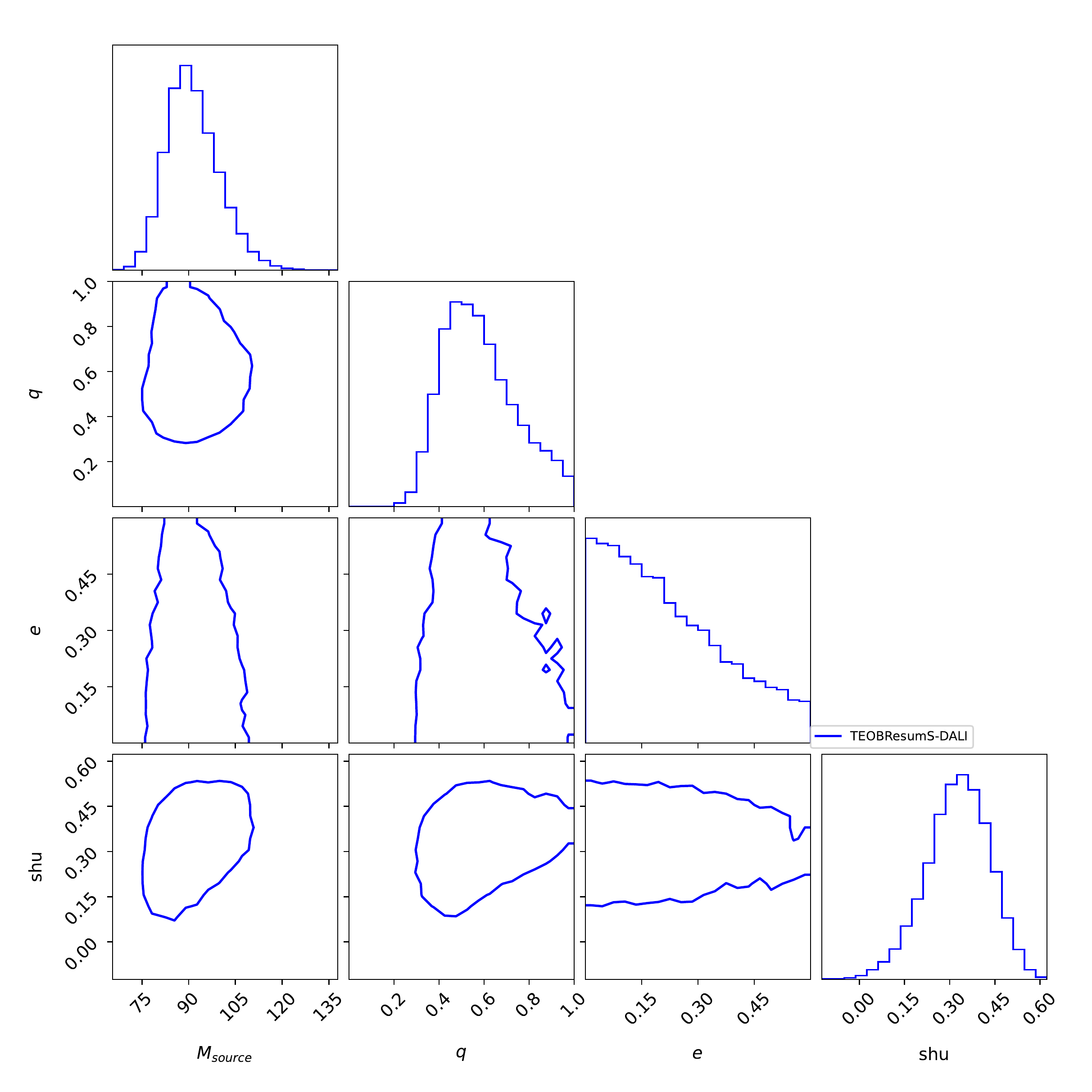}
    }
    \subfloat{
        \includegraphics[width=.5\textwidth]{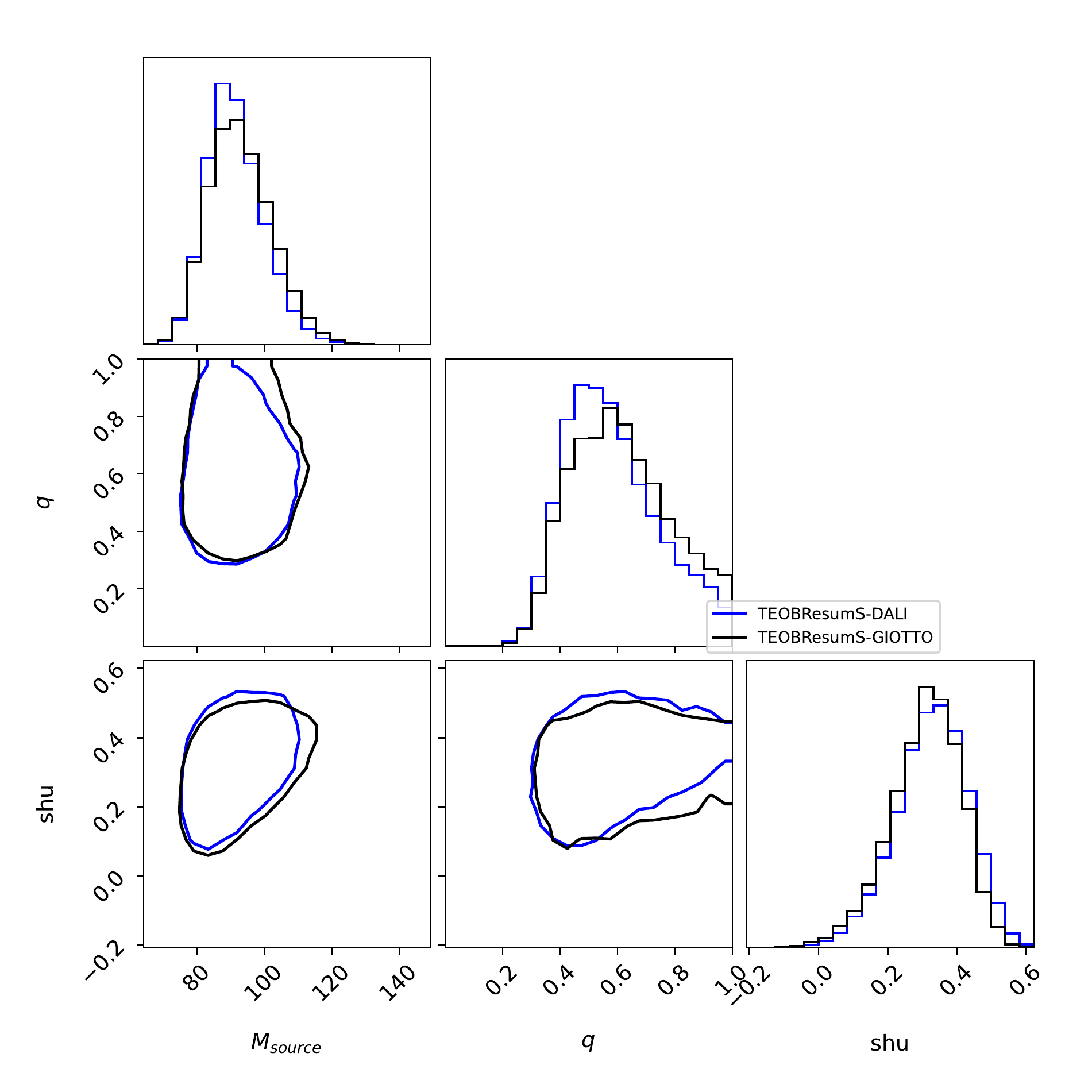}
    }
    \caption{
        \label{fig:GW190620}
        \textbf{GW190620}: Corner plots of 2D and 1D marginal posteriors of $M_{\rm source},q,e,S_{\rm hu}$ using TEOBResumS-GIOTTO in black and TEOBResumS-DALI in blue. \textit{The left panel} shows only the analysis using TEOBResumS-DALI and includes eccentricity in the corner plot along with masses and spin parameters. \textit{The right panel} includes both the TEOBResumS-DALI and TEOBResumS-GIOTTO analyses including only the masses and spin parameters. The contours represent the 90\% confidence intervals for each joint distribution. When eccentricity is included, it slightly shifts the posterior of the masses and spins.
        }
\end{figure*}
\begin{figure*}
    \subfloat{
        \includegraphics[width=.5\textwidth]{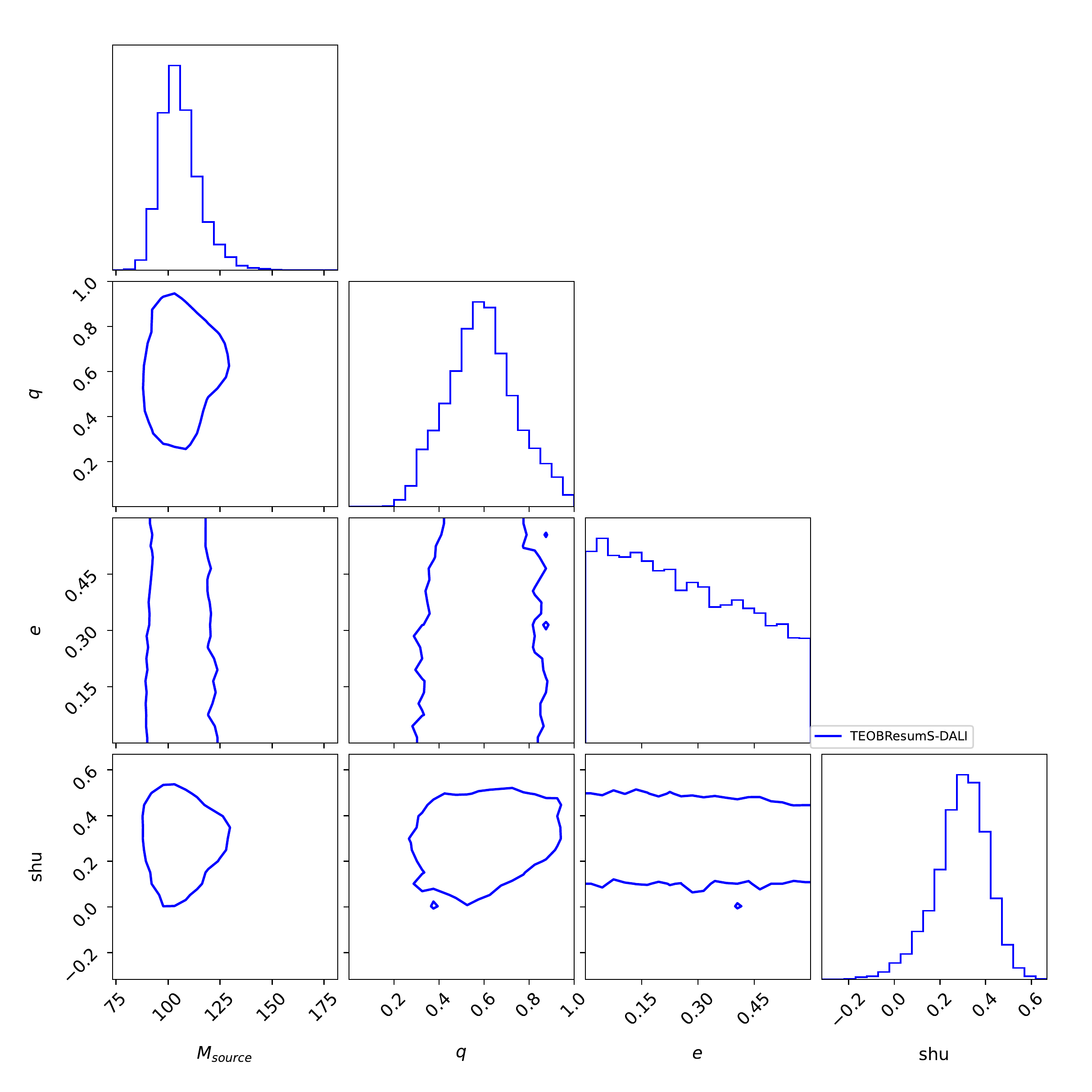}
    }
    \subfloat{
        \includegraphics[width=.5\textwidth]{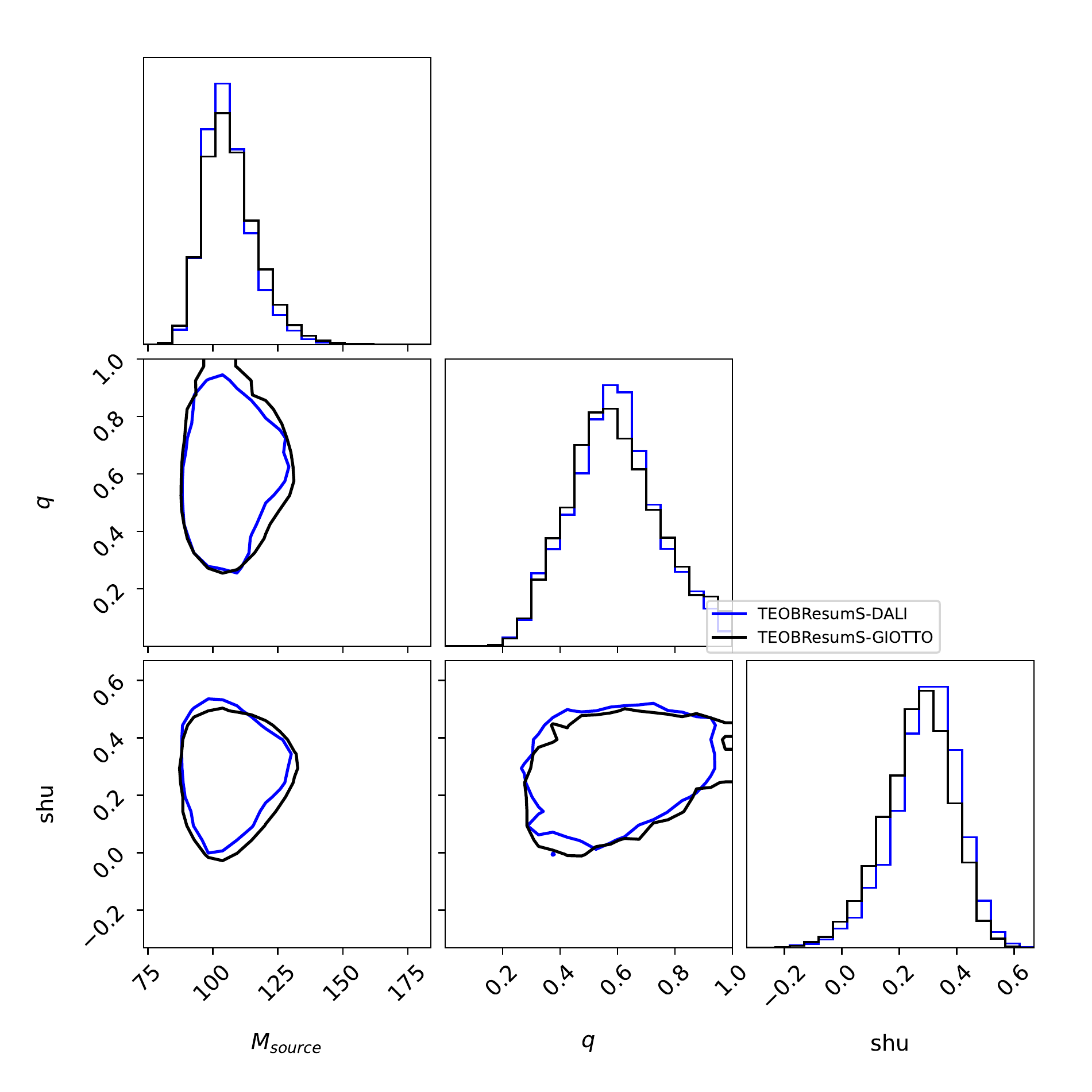}
    }
    \caption{
        \label{fig:GW190706}
        \textbf{GW190706}: Corner plots of 2D and 1D marginal posteriors of $M_{\rm source},q,e,S_{\rm hu}$ using TEOBResumS-GIOTTO in black and TEOBResumS-DALI in blue. \textit{The left panel} shows only the analysis using TEOBResumS-DALI and includes eccentricity in the corner plot along with masses and spin parameters. \textit{The right panel} includes both the TEOBResumS-DALI and TEOBResumS-GIOTTO analyses including only the masses and spin parameters. The contours represent the 90\% confidence intervals for each joint distribution. When eccentricity is included, it slightly shifts the posterior of the masses and spins.
        }
\end{figure*}
\begin{figure*}
    \subfloat{
        \includegraphics[width=.5\textwidth]{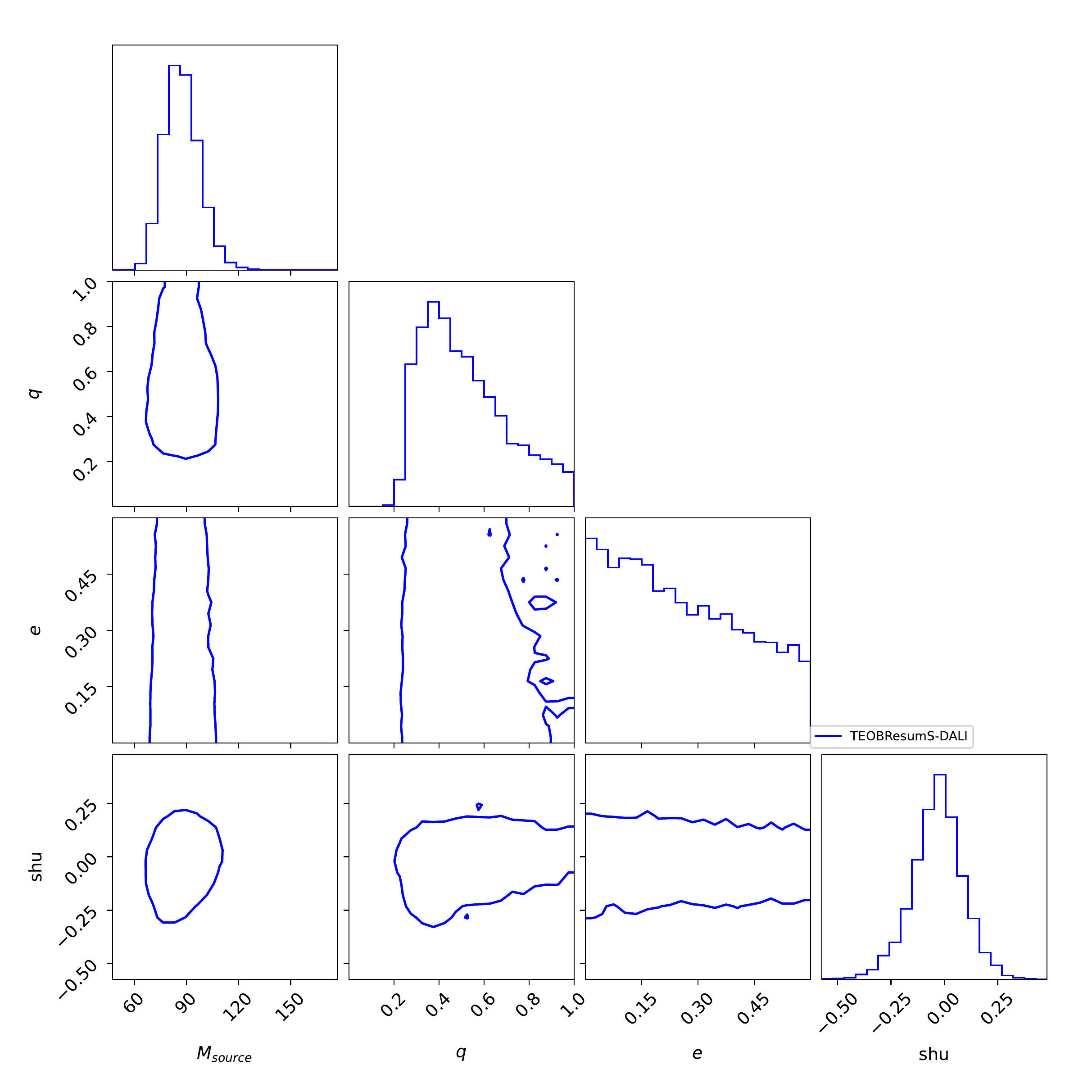}
    }
    \subfloat{
        \includegraphics[width=.5\textwidth]{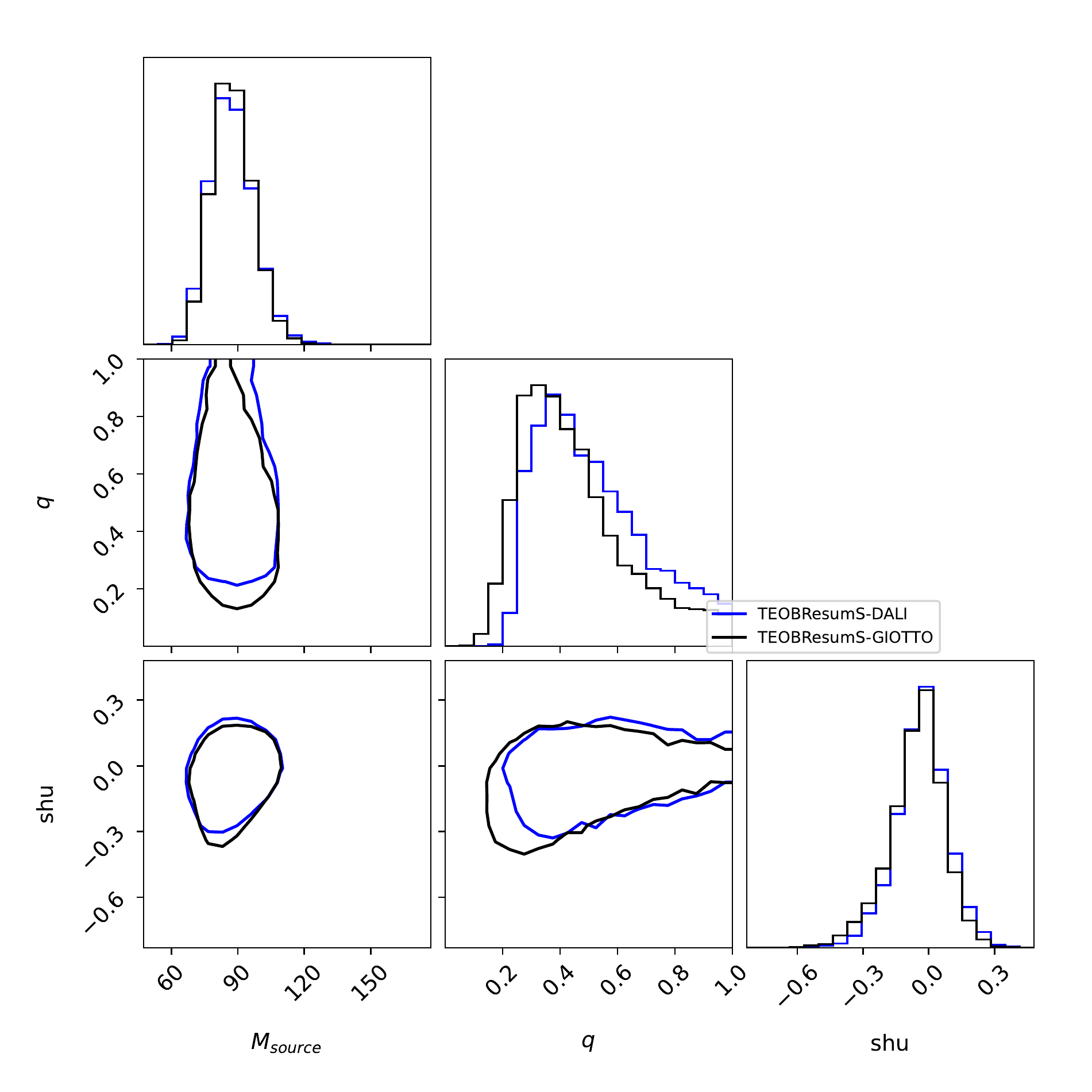}
    }
    \caption{
        \label{fig:GW190929}
        \textbf{GW190929}: Corner plots of 2D and 1D marginal posteriors of $M_{\rm source},q,e,S_{\rm hu}$ using TEOBResumS-GIOTTO in black and TEOBResumS-DALI in blue. \textit{The left panel} shows only the analysis using TEOBResumS-DALI and includes eccentricity in the corner plot along with masses and spin parameters. \textit{The right panel} includes both the TEOBResumS-DALI and TEOBResumS-GIOTTO analyses including only the masses and spin parameters. The contours represent the 90\% confidence intervals for each joint distribution. When eccentricity is included, it shifts the posterior of the mass ratio.
        }
\end{figure*}
\subsection{Spins}
\label{subsec:spins}
Similar to Section \ref{subsec:masses}, we can also directly compare the analyses to examine how the inclusion of eccentricity affects the recovery of the spin parameters. For our Figures, we use the spin parameter $S_{\rm hu}$ (\cite{2018PhRvD..97h4002H}) and is defined as:
\begin{equation}
    M^2S_{\rm hu}=\left((1+\frac{1}{2q})\Vec{S}_1+(1+\frac{1}{2}q)\Vec{S}_2\right)\cdot\hat{L}.
\end{equation}
The $S_{\rm hu}$ parameter is similar to the widely used $\chi_{\rm eff}$; however, $S_{\rm hu}$ describes the leading order effect of hangup on the full NR waveforms.

The inferred spin and mass-spin distributions are frequently quite consistent in shape and support when using a model including eccentricity with only GW150914 being the outlier having a $\epsilon_{S_{\rm hu}}=0.834$. Interestingly, the inclusion of eccentricity shifts the spin parameter posterior to more positive values in the case of GW150914. Unsurprisingly few finely-tuned accidents occur, so binaries with notable shifts in inferred masses usually also have notable shifts in inferred spin, and vice-versa. 

\subsection{Eccentricity}
\label{subsec:eccentricity}
Despite the impact of eccentricity on other parameters, our inferred eccentricity distributions are frequently quite modest. For example, the inferred distribution of eccentricity for GW150914 strongly supports a non-eccentric origin. Even for GW190521, where the inferred posterior eccentricity peaks near $e \simeq \,0.2$, the posterior distribution contains significant support for $e\simeq 0$. A Bayes factor analysis performed both by direct integration and by the Savage-Dickey ratio suggests these two events are most likely non-eccentric, with Bayes' factors of 0.153 and 0.174, respectively. In fact, as Table \ref{tab:ecc} shows, none of the events strongly prefer the eccentric hypothesis over the non-eccentric hypothesis. Our conclusions about these events are reasonably consistent with prior studies using \emph{non-precessing} binaries, which generally find at best modest evidence for eccentricity when precession is not allowed. While on the surface our numerical Bayes factors for these events are in notable tension with the results of Romero-Shaw et al,  we highlight the substantial systematic and methodological differences associated with our approach using a different eccentric model with a wider prior range and a different starting frequency. While a direct apples-to-apples comparison between these results would need a conversion between the two models' definition of eccentricity, a closer comparison would require similar run settings. Such a conversion has been investigated in \cite{2022ApJ...936..172K}, who developed a framework to translate between the different definitions of eccentricity in the eccentric waveform models SEOBNRE and TEOBResumSGeneral.

While not fully understood, it is expected that eccentricity could mimic precession. To ensure the evidences presented here are not mistaken for evidences of misaligned spins, we conduct analyses of GW150914, GW190521, GW190620, GW190706, and GW190929 using the precessing waveform model TEOBResumSP-GIOTTO; this allows us to perform a comparison between the eccentric and precession hypotheses. The posterior distributions obtained with all three models, along with a discussion of the results from the precession analyses, can be found in Appendix \ref{app:precession}.

In addition to the eccentric vs. non-eccentric Bayes' factors $\text{BF}_{E}$, Table \ref{tab:ecc} contains the Bayes' factors $\text{BF}_{P}$ of the quasi-circular, spin-precessing hypothesis, calculated with the waveform model TEOBResumSP-GIOTTO, against the spin-aligned hypothesis, calculated with the waveform model TEOBResumS-GIOTTO, for GW150914; GW190521; GW190620; GW190706; and GW190929. We have also included the relative Bayes' factors $\text{BF}_{E/P}$ of the eccentric hypothesis, calculated with the waveform model TEOBResumS-DALI, against the precessing hypothesis. Our results for $\text{BF}_{P}$ indicate a marginal preference for the spin-precession hypothesis for GW150914, GW190706, and GW190929; furthermore, our results for $\text{BF}_{E/P}$ indicate a marginal preference for the spin-precession hypothesis for all of the events. However, none of the events exhibit a strong preference for either hypothesis, so we cannot conclusively state whether any of these events are eccentric, precessing, or manifest both eccentricity and spin-precession.

\begin{figure*}
    \subfloat{
        \includegraphics[width=.49\textwidth]{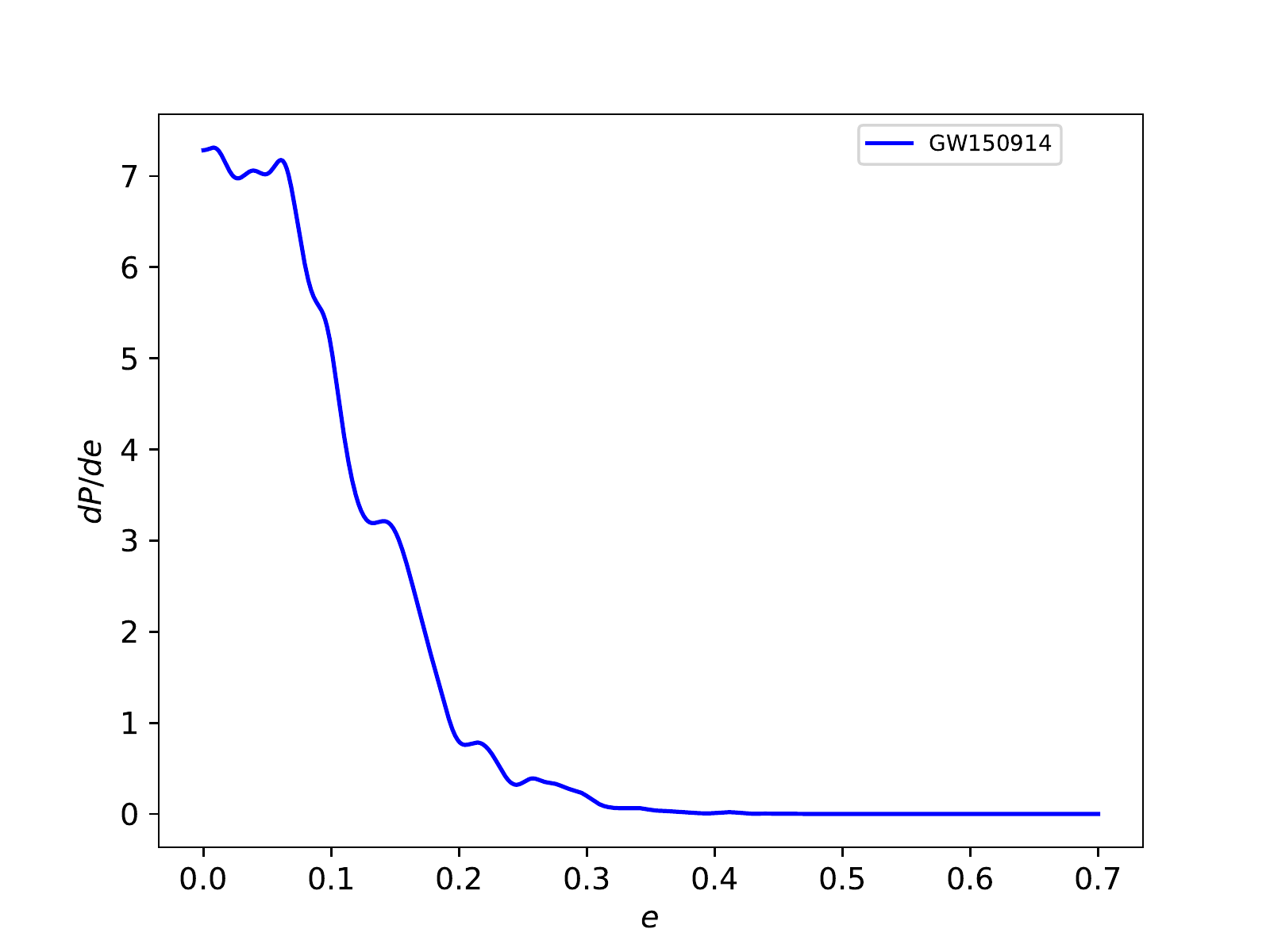}
    }
    \subfloat{
        \includegraphics[width=.49\textwidth]{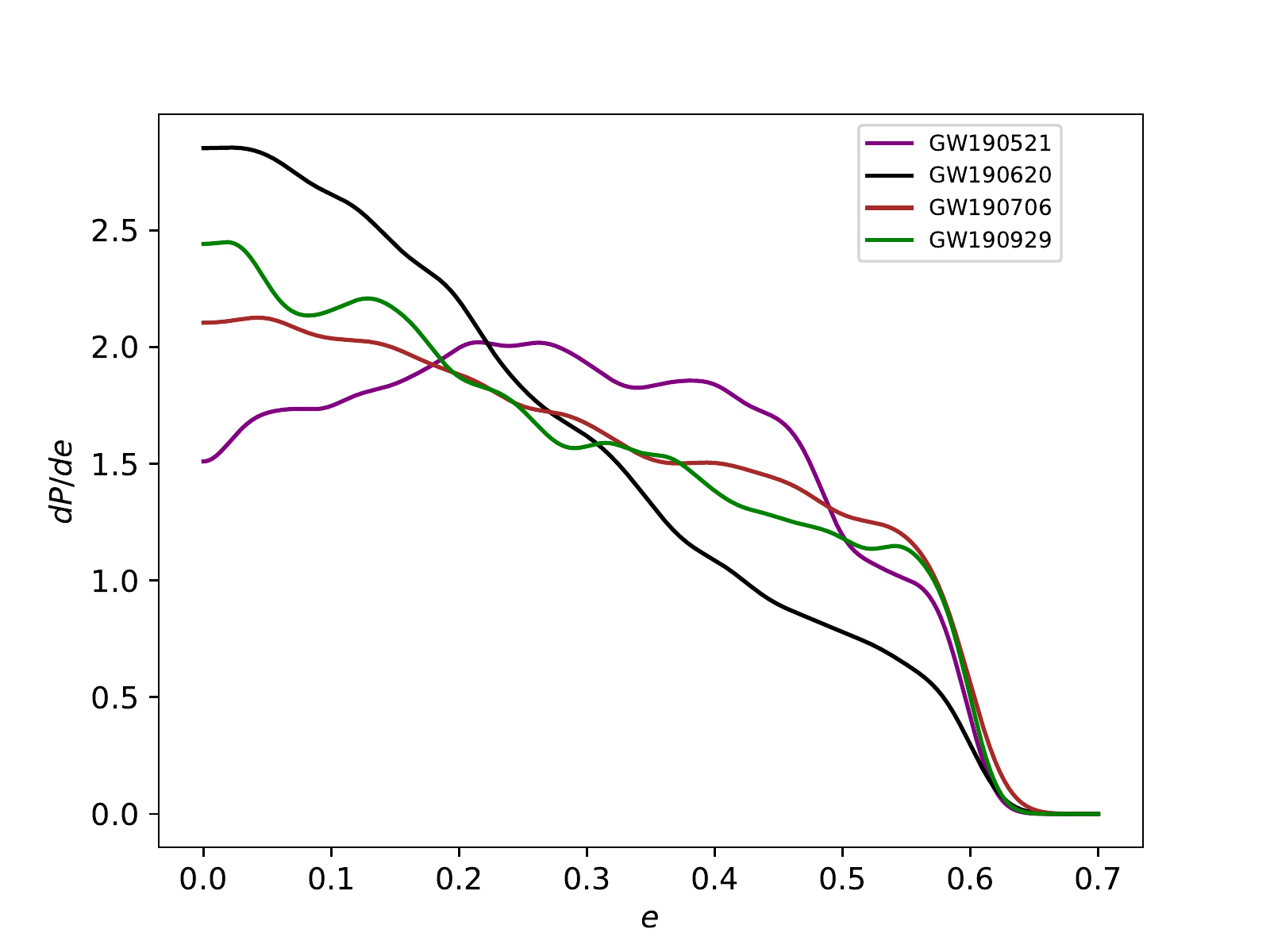}
    }
    \caption{
        \label{fig:ecc}
        \textbf{1D marginal distribution of eccentricity}: Each distribution is the 1D marginal distribution of eccentricity for each event: GW150914, GW190521, GW190620, GW190706, GW190929 (this is the same as one of the histograms in the left panels of Figures \ref{fig:GW150914}-\ref{fig:GW190929}). The left panel is only of GW150914 due to the significantly different scaling between this and the other distributions. See Table \ref{tab:ecc} for the Bayes' Factors comparing eccentric vs non-eccentric.
        }
\end{figure*}
\begin{table}
\caption{\label{tab:ecc}%
Bayes' factors for each event comparing evidences of eccentric vs. non-eccentric ($\text{BF}_{E}$), precessing vs. non-eccentric ($\text{BF}_{P}$), and eccentric vs. precessing ($\text{BF}_{E/P}$). 
}
\begin{tabular}{ m{7em} m{5em} m{5em} m{5em}}
\toprule
    \textrm{Event} &
    \textrm{$\text{BF}_{E}$} &
    \textrm{$\text{BF}_{P}$} &
    \textrm{$\text{BF}_{E/P}$}\\
\colrule
    GW150914 & 0.153 & 1.626 & 0.094 \\
    GW190521 & 0.174 & 0.492 & 0.357 \\
    GW190620 & 0.134 & 0.539 & 0.248 \\
    GW190706 & 0.326 & 1.088 & 0.299 \\
    GW190929 & 0.616 & 1.235 & 0.498 \\
\botrule
\end{tabular}
\end{table} \break
\section{Conclusions}
\label{sec:conclusions}
In this work, we presented a state-of-the-art  model-based assessment of the presence and impact of eccentricity in several promising BBH events, based on the TEOBResumSGeneral model, including higher-order modes.  We compare two analyses with the same model that include and exclude eccentricity. 
While we do see some shifts in the posterior when including eccentricity, we do not find evidence in favor of eccentric dynamics for any of the events presented, instead favoring a quasi-circular result.%
%
 Our results are in tension numerically with previously-presented results for several of these events. Specifically we do not find evidence of eccentricity for either GW190521 or GW190620; however, numerical differences are most likely due to the differences in prior range and starting frequencies, and it could also reflect systematics, as we adopted different waveforms and algorithms.

For example, \cite{2022NatAs...6..344G} found higher likelihoods for eccentric, precessing systems compared to quasi-circular, precessing systems for GW190521, using direct comparison to precessing numerical relativity simulations; by contrast, our analysis assumes spin-orbit alignment and finds the full parameter posterior.
Likewise, \cite{2021arXiv210605575G} argued that GW190521 is highly eccentric based on an analysis with TEOBResumSGeneral, excluding higher-order modes, starting from an initially unbound configuration at extremely low initial frequencies; by contrast, our analysis exclusively uses only the bound,  later-time evolution (and thus higher starting frequency and corresponding lower eccentricity) and the higher-order-mode form of that model.
Similarly, \cite{2021ApJ...921L..31R} found evidence of eccentricity for GW190620 and GW190521 based off analyses that used a re-weighting technique with the eccentric model SEOBNRE, a model that excludes higher-order modes, using a different configuration with different priors and initial frequencies; by contrast, our analysis again uses a starting frequency of 5~Hz and higher-order-mode form of a different model as well as a larger uniform-in-e range. To do a fair comparison, one would need to use the same settings of one of the above analyses.

In future work, we plan to analyze the rest of the public events available from the LVK \citep{2023ApJS..267...29A,2021SoftX..1300658A}. We plan to include the same three configurations as in this paper: the non-eccentric, non-precessing case; the eccentric, non-precessing case; and the non-eccentric, precessing case to make sure any evidence of eccentricity is not mixing up evidence of precession.

%

\acknowledgements
The authors thank Juan Caldron Bustillo and Aaron Zimmerman for useful feedback and conversations. We are thankful to the waveform developers of TEOBResumSGeneral, specifically Alessandro Nagar and Sebastiano Bernuzzi for their help with implementing their model. The git hash of TEOBResumSGeneral used was bc4fcd0f0d97f9f0351e11c2e880ab0a6422ac20 for non-eccentric and 39e6d7723dacb23220ff5372e29756e5f94cb004 for eccentric I.B., V.G., and S.B. acknowledges the support of the National Science Foundation under grants PHY-1911796 and PHY-2110060, and the Alfred P. Sloan Foundation. HLI, JL, AJ, RN, DS, and RV thanks NSF PHY-2114581, PHY-2207780 and XSEDE TG-PHY120016. G.V acknowledges the support of the
National Science Foundation under grant PHY-2207728.
ROS is supported by NSF PHY-2012057, PHY-1912632, 
and AST-1909534.
 This material is based upon work supported by NSF's LIGO Laboratory which is a major facility fully funded by the National Science Foundation. This research has made use of data or software obtained from the Gravitational Wave Open Science Center (gw-openscience.org), a service of LIGO Laboratory, the LIGO Scientific Collaboration, the Virgo Collaboration, and KAGRA. LIGO Laboratory and Advanced LIGO are funded by the United States National Science Foundation (NSF) as well as the Science and Technology Facilities Council (STFC) of the United Kingdom, the Max-Planck-Society (MPS), and the State of Niedersachsen/Germany for support of the construction of Advanced LIGO and construction and operation of the GEO600 detector. Additional support for Advanced LIGO was provided by the Australian Research Council. Virgo is funded, through the European Gravitational Observatory (EGO), by the French Centre National de Recherche Scientifique (CNRS), the Italian Istituto Nazionale di Fisica Nucleare (INFN) and the Dutch Nikhef, with contributions by institutions from Belgium, Germany, Greece, Hungary, Ireland, Japan, Monaco, Poland, Portugal, Spain. The construction and operation of KAGRA are funded by Ministry of Education, Culture, Sports, Science and Technology (MEXT), and Japan Society for the Promotion of Science (JSPS), National Research Foundation (NRF) and Ministry of Science and ICT (MSIT) in Korea, Academia Sinica (AS) and the Ministry of Science and Technology (MoST) in Taiwan.

We are grateful for computational resources provided by the Leonard E Parker Center for Gravitation, Cosmology and Astrophysics at the University of Wisconsin-Milwaukee. We acknowledge the use of IUCAA LDG cluster Sarathi for the 
computational/numerical work. We are also grateful for the computational resources provided by California Institute of Technology at Pasadena, California as well as the LIGO Livingston Observatory.

\bibliography{Refs}

\appendix
\section{Waveform model}
\label{app:waveform model}
\begin{figure}
    \subfloat{
        \includegraphics[width=.49\textwidth]{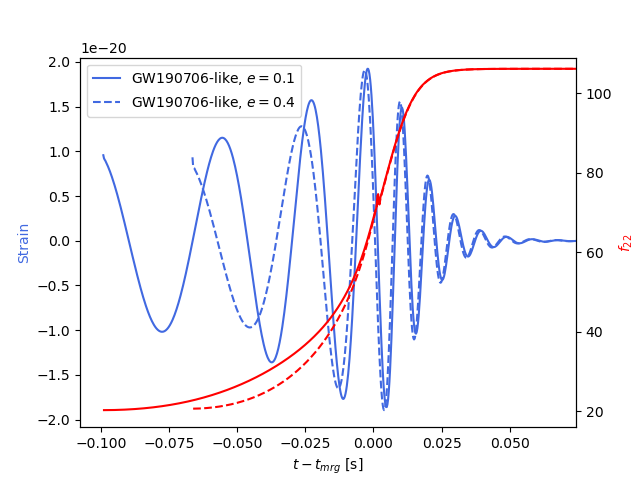}
    }
    \subfloat{
        \includegraphics[width=.49\textwidth]{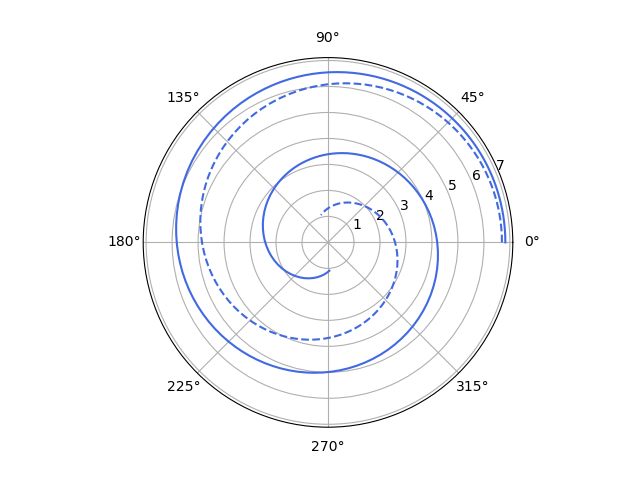}
    }
    \caption{
        \label{fig:wfexample}
        \textbf{GW190706 waveform and dynamics}: Waveforms (left) and EOB dynamics (right) corresponding to the two eccentricity peaks of the GW190706 analysis ($e_0=0.1$, $e_0=0.4$). 
        Because of the low radial separation at which the evolution of the system is started, the waveforms emitted by the two systems look remarkably similar, in spite of the rather different values of input eccentricity.
}
\end{figure}

The TEOBResumS-DALI model employed in the main text can generate waveforms over a large portion of the parameter space, 
including systems with large eccentricities ($e \sim 0.9$).
However, extreme care should be applied when such regions are explored for systems with large total detector frame mass. 
Indeed, if $M$ is large and the initial (reference) frequency is high, the initial separation $r_0$
at which the EOB model will begin to evolve the dynamics of the system may be close to the last stable orbit (LSO) of the binary.
In this scenario, bodies evolving along eccentric orbits may plunge almost immediately, completing very few orbital cycles 
and -- as a consequence -- emitting very similar waveforms irrespectively of the starting input eccentricity.
A clear example of this can be seen in Fig.~\ref{fig:wfexample}, where we display the waveforms and EOB dynamics obtained with the parameters corresponding to the two posterior peaks of the GW190706 analysis.
%
To avoid a multi-modal eccentricity estimation that is a direct  consequence of evolving the system with separation close to its LSO, we lowered the initial reference frequency in our analysis from 18~Hz to 5~Hz. 

\section{Precession analysis} 
\label{app:precession}
\begin{figure}
    \centering
    \includegraphics[width=.5\textwidth]{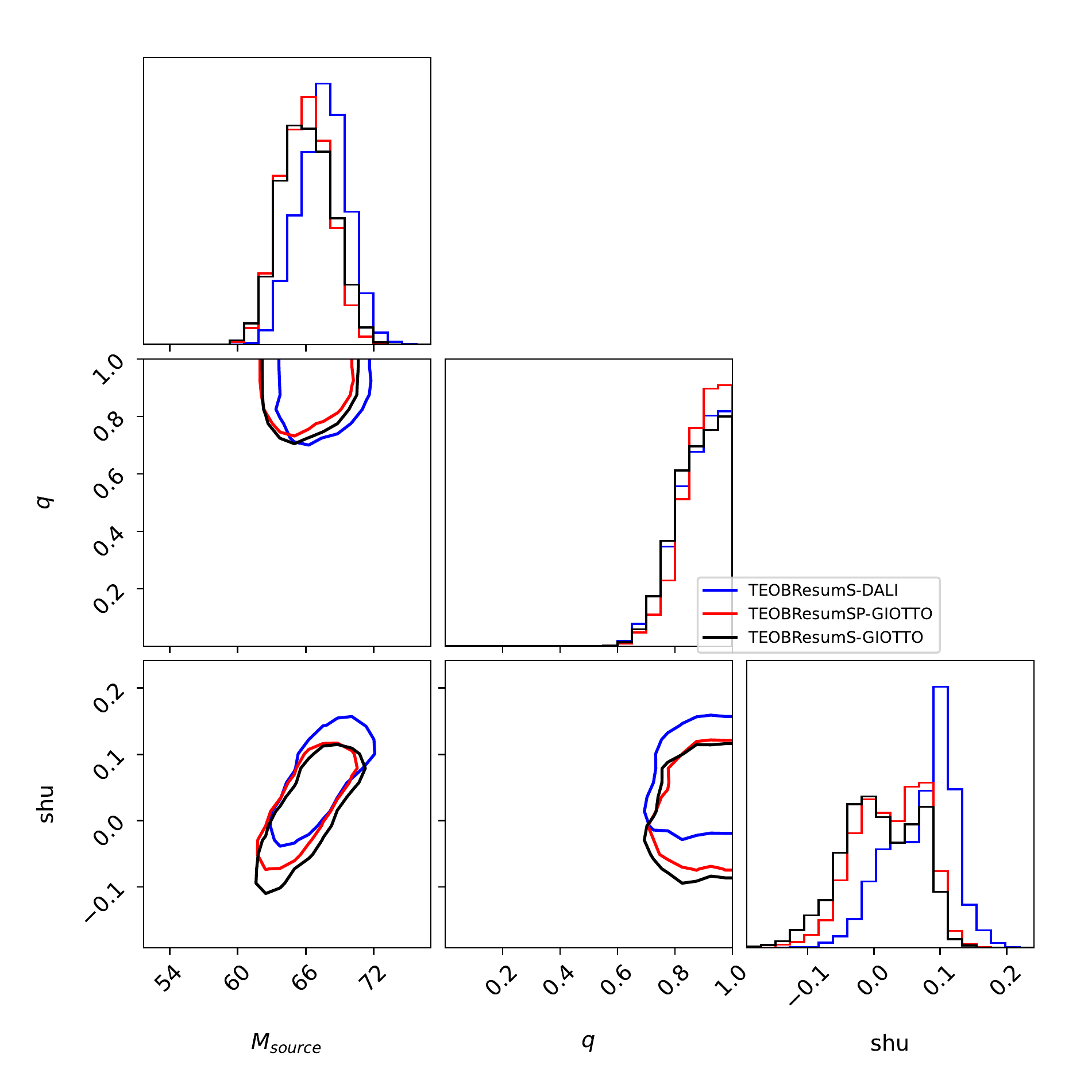}
    \caption{\textbf{GW150914 (all analyses)}: A corner plot of 2D and 1D marginal posteriors of $M_{\rm source},q,S_{\rm hu}$ using TEOBResumS-GIOTTO in black, TEOBResumSP-GIOTTO in red, and TEOBResumS-DALI in blue. The contours represent the 90\% confidence intervals for each joint distribution. When eccentricity is included, it shifts the posterior of the masses and spins.}
    \label{fig:GW150914_combined}
\end{figure}

\begin{figure}
    \centering
    \includegraphics[width=.5\textwidth]{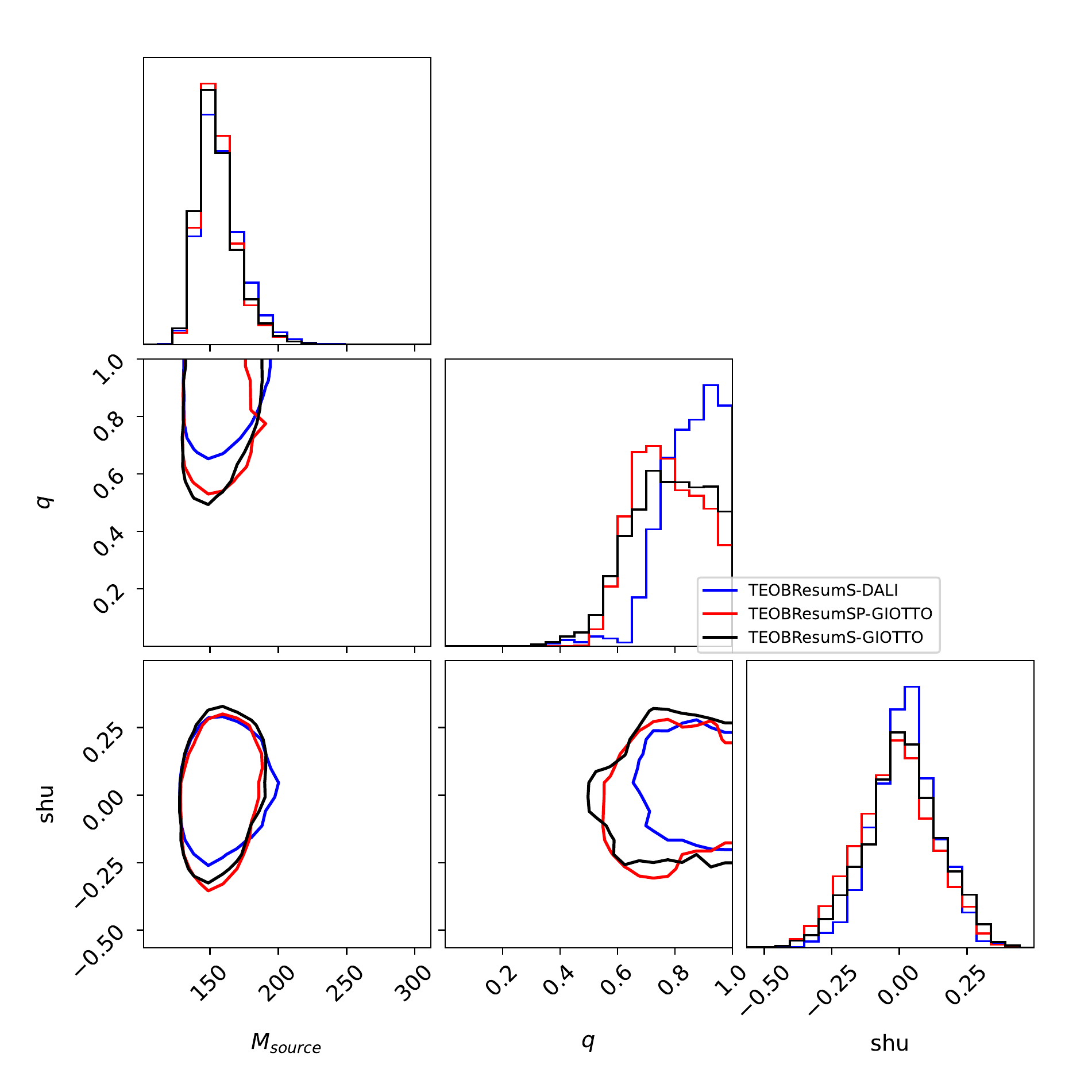}
    \caption{\textbf{GW190521 (all analyses)}: A corner plot of 2D and 1D marginal posteriors of $M_{\rm source},q,S_{\rm hu}$ using TEOBResumS-GIOTTO in black, TEOBResumSP-GIOTTO in red, and TEOBResumS-DALI in blue. The contours represent the 90\% confidence intervals for each joint distribution. When eccentricity is included, it shifts the posterior of the mass ratio.}
    \label{fig:GW190521_combined}
\end{figure}

\begin{figure}
    \centering
    \includegraphics[width=.5\textwidth]{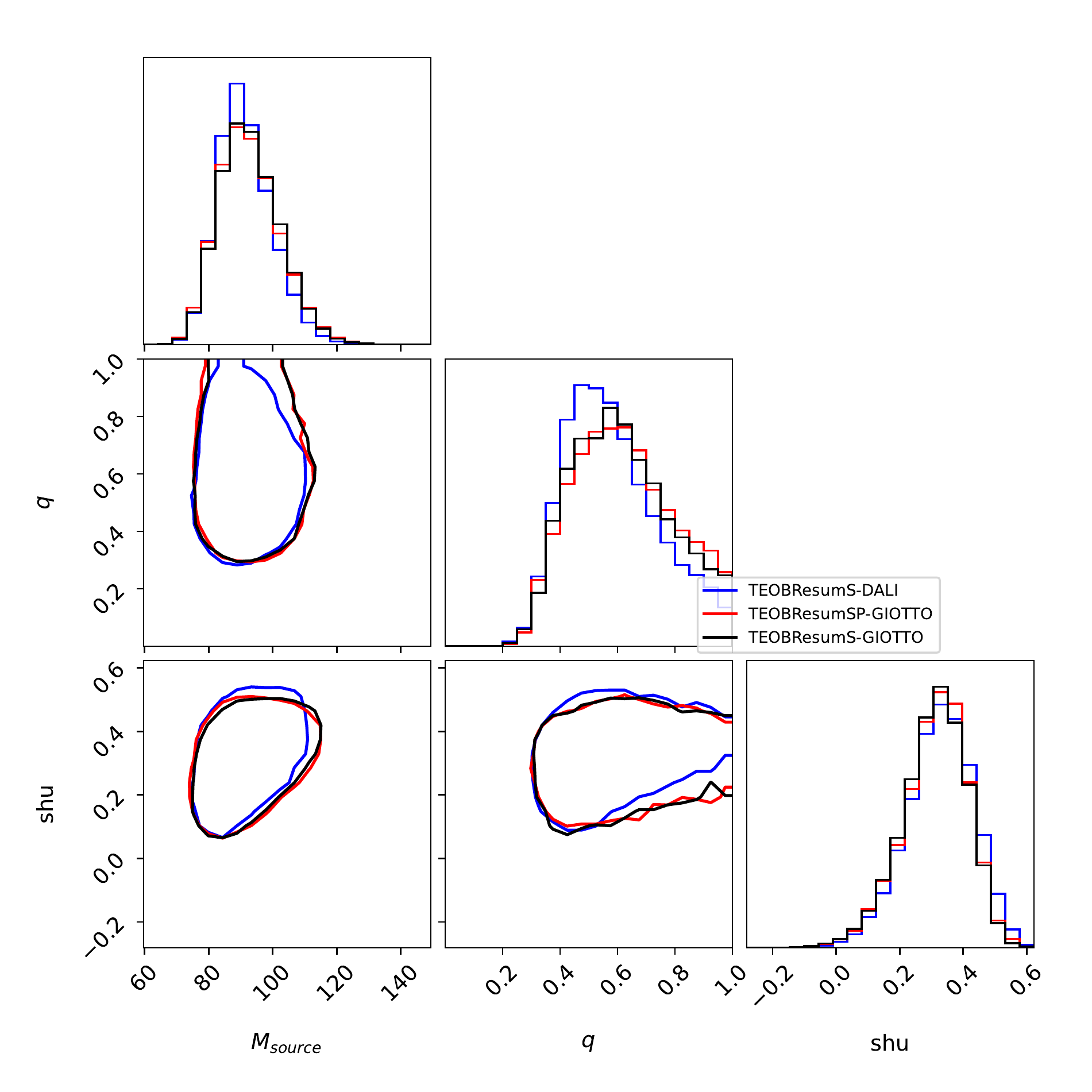}
    \caption{\textbf{GW190620 (all analyses)}: A corner plot of 2D and 1D marginal posteriors of $M_{\rm source},q,S_{\rm hu}$ using TEOBResumS-GIOTTO in black, TEOBResumSP-GIOTTO in red, and TEOBResumS-DALI in blue. The contours represent the 90\% confidence intervals for each joint distribution. When eccentricity is included, it slightly shifts the posterior of the masses and spins.}
    \label{fig:GW190620_combined}
\end{figure}

\begin{figure}
    \centering
    \includegraphics[width=.5\textwidth]{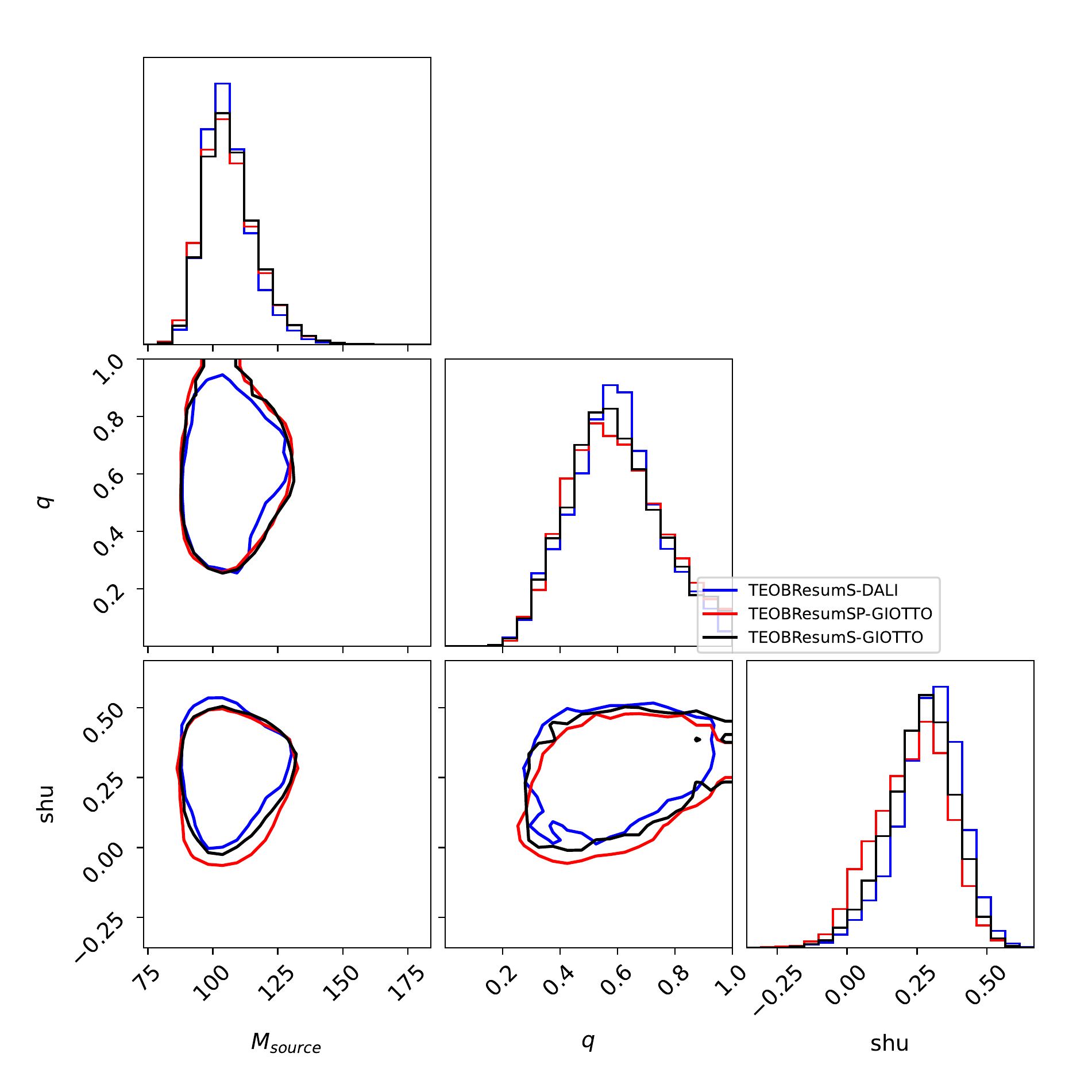}
    \caption{\textbf{GW190706 (all analyses)}: A corner plot of 2D and 1D marginal posteriors of $M_{\rm source},q,S_{\rm hu}$ using TEOBResumS-GIOTTO in black, TEOBResumSP-GIOTTO in red, and TEOBResumS-DALI in blue. The contours represent the 90\% confidence intervals for each joint distribution. When eccentricity is included, it slightly shifts the posterior of the masses and spins.}
    \label{fig:GW190706_combined}
\end{figure}

\begin{figure}
    \centering
    \includegraphics[width=.5\textwidth]{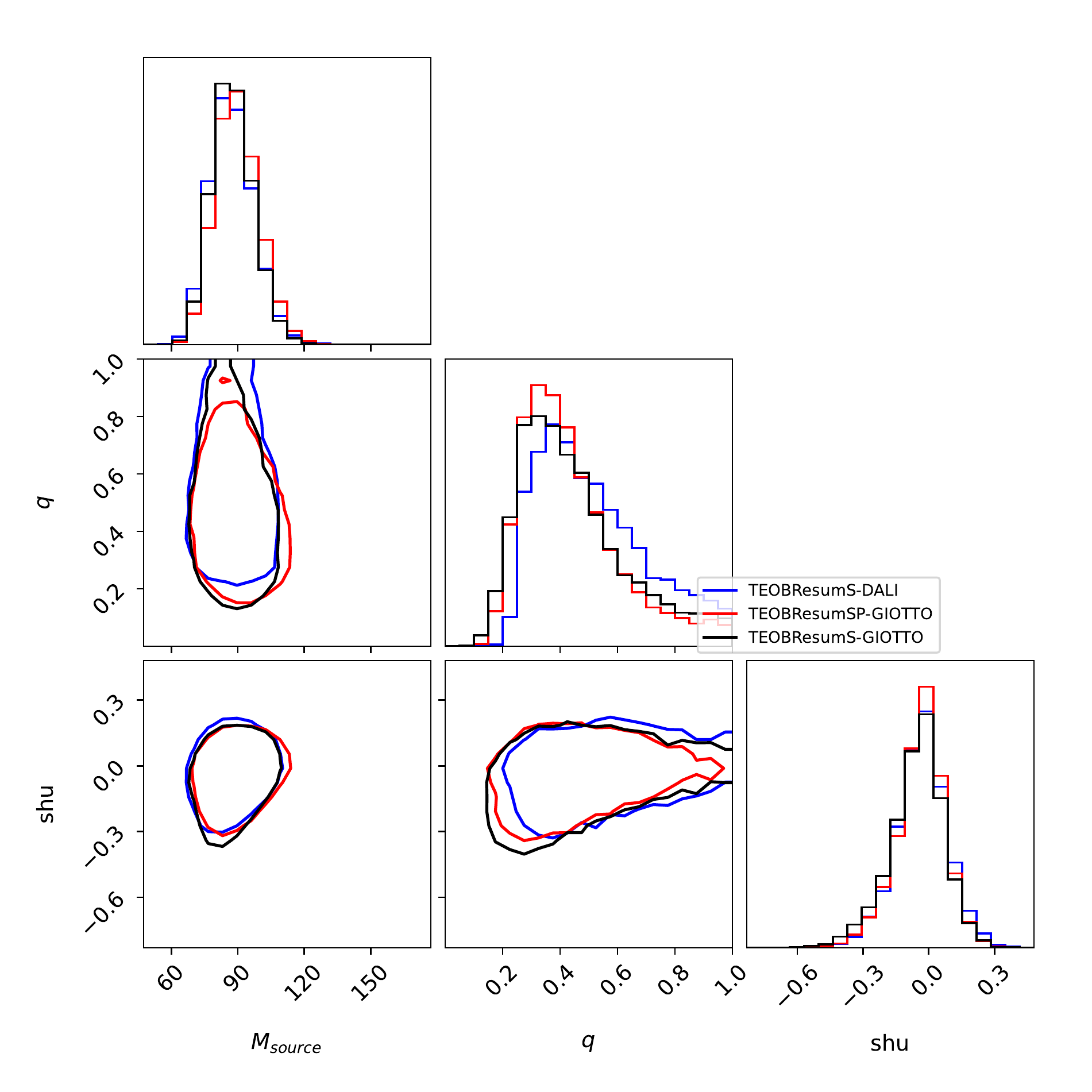}
    \caption{\textbf{GW190929 (all analyses)}: A corner plot of 2D and 1D marginal posteriors of $M_{\rm source},q,S_{\rm hu}$ using TEOBResumS-GIOTTO in black, TEOBResumSP-GIOTTO in red, and TEOBResumS-DALI in blue. The contours represent the 90\% confidence intervals for each joint distribution. When eccentricity is included, it shifts the posterior of the mass ratio.}
    \label{fig:GW190929_combined}
\end{figure}

We perform quasi-circular, spin-precessing analyses of the events GW150914, GW190521, GW190620, GW190706, and GW190929, using the precessing waveform model TEOBResumSP-GIOTTO. For the precessing spins, we assume the spin vectors to be isotropic on the sphere and uniform in spin magnitude $\chi_{i,\{x,y,z\}}\in[-0.99,0.99]$. Figures \ref{fig:GW150914_combined}, \ref{fig:GW190521_combined}, \ref{fig:GW190620_combined}, \ref{fig:GW190706_combined}, and \ref{fig:GW190929_combined} show posterior distributions for mass and spin parameters calculated from all three analyses for our five selected events. Across all five events, the posteriors for the mass and spin parameters using the spin-precessing model are fairly consistent in shape and support when compared to the posterior distributions generated using the aligned spin model. 


\end{document}